\documentclass[twocolumn,showpacs,preprintnumbers,amsmath,amssymb]{revtex4}
\usepackage{graphicx}
\usepackage{dcolumn}
\usepackage{bm}
\begin{document}
\title{A Renormalization Group Procedure for Fiber Bundle Models}
\author{Srutarshi Pradhan\email{srutarshi.pradhan@ntnu.no}}
\affiliation{PoreLab, Department of Physics, Norwegian University of
Science and Technology, NO--7491 Trondheim, Norway.}
\author{Alex Hansen\email{alex.hansen@ntnu.no}}
\affiliation{PoreLab, Department of Physics, Norwegian University of
Science and Technology, NO--7491 Trondheim, Norway.}
\author{Purusattam Ray\email{ray@icms.ac.in}}
\affiliation{The Institute of Mathematical Sciences, CIT Campus,
Taramani,Chennai 600 113, India.}
\affiliation{Homi Bhabha National Institute, Training School Complex, Anushakti Nagar, Mumbai 400094, India.}
\date{\today {}}
\begin{abstract}
We introduce two versions of a renormalization group scheme for the equal
load sharing fiber bundle model.  The renormalization group is based on
formulating the fiber bundle model in the language of damage mechanics.
A central concept is the work performed on the fiber bundle to produce a
given damage. The renormalization group conserves this work.  In the first 
version of the renormalization group, we take advantage of ordering the strength
of the individual fibers. This procedure, which is the simpler one, gives 
EXACT results -but cannot be generalized to other fiber bundle models 
such as the local load
sharing one.  The second renormalization group scheme based on the 
physical location of the individual fibers may be generalized to other 
fiber bundle models. 
\end{abstract}
\maketitle

\section{Introduction} 
\label{intro}

In an age where computer modeling of fracture and material breakdown 
is reaching a stage where the systems one may study span from the atomistic 
level to the continuum level in a single go \cite{b07,b08}, one may 
wonder what is the use of simplified models. The fact is that such 
modeling is more important than ever.  The experimental approach tells 
us what Nature dictates the systems to do.  The computational approach 
tells us what would happen if we, and not Nature, made the rules.  With 
the computational approach we are able to know exactly what every single 
atom in the material is doing.  However, this is not equivalent to 
{\it understanding\/} what is happening. For this, we need to find the 
underlying principles, and this is where the need for simplified models 
come in.  

Looking back in history, the study of equilibrium critical phenomena became
``well understood" in the late seventies.   Central to conquest of this 
field was the Ising spin model \cite{wk74,f17}.  It is hard to imagine
how the deep understanding of the nature of critical phenomena could have
evolved without the guidance of this model and all its more complex
relatives. 

There are similarities between equilibrium phenomena and fracture, but
also large differences.  The similarities come from the development of 
long-range correlations as the fracture process proceeds in the same
way as such correlations develop when approaching a critical point.  On
the other hand, whereas parameters need to be adjusted to approach 
criticality in equilibrium systems, in fracture the system approaches
this state without the tuning of parameters.  The correlations that
develop during the fracture process stem from the  way the stress
field develops.  However, they also reflect themselves in e.g.\
the spatial correlations in the post mortem fracture surfaces 
\cite{bb11,brc15}.

The fiber bundle model \cite{p26,d45} is a model that plays
somewhat the same role with respect to fracture phenomena as the Ising
model plays with respect to equilibrium critical phenomena \cite{phc10,hhp15}.
In its simplest form, the {\it Equal Load Sharing\/} (ELS) version,
it consists of $N$ parallel fibers of length $L$ placed between two 
parallel stiff clamps a distance $L+\Delta$ apart.  
Each fiber responds linearly with a force $f$ to the load $\Delta$,
\begin{equation}
\label{eq1-0}
f=\kappa \Delta\;,
\end{equation}
where $\kappa$ is the spring constant.  $\kappa$ is the same for all fibers.
Each fiber has a load threshold $t$ assigned to it. The load thresholds
are drawn randomly from a probability density $p(t)$. If the load $\Delta$ 
exceeds this threshold, the fiber fails irreversibly.  The total load on
the fiber bundle is 
\begin{equation}
\label{eq2-0}
F=\kappa(N-n)\Delta\;
\end{equation}
when $n$ fibers have failed. This means that all thresholds $t\le\Delta$
have failed.

It is the aim of this paper to construct a real-space renormalization group 
scheme \cite{h82} for the ELS fiber bundle model. We hope, the renormalization 
group scheme is generalizable to more
complex fiber bundle models such as the {\it Local Load Sharing\/} (LLS)
fiber bundle model \cite{hp91} or the {\it Soft Clamp\/} (SC) fiber bundle
model \cite{bhs02}.  

Our goal is to construct a mapping from a fiber bundle containing $N$ fibers
to a fiber bundle containing $N'=N/2$ fibers in such a way that the variables
describing the entire fiber bundle, such as $F$ and $\Delta$ remain unaltered.
This we do by replacing the fibers characterized by a spring constant $\kappa$
and threshold distribution $p(t)$ by a new set of fibers characterized by 
a spring constant $\kappa'$ and a threshold distribution $p'(t)$. 

In order to construct the real space renormalization group, it is necessary
to formulate the ELS fiber bundle model within {\it damage 
mechanics\/} \cite{k89,a15,b15}. We present in Section \ref{cdm}
a new formulation of the ELS fiber bundle model within such a framework
tailored for the renormalization group formulation to be presented.

It is an important feature of the ELS fiber bundle model that it is
infinite dimensional.  That is, all fibers interact with all other
fibers in exactly the same way.  This is in contrast to e.g.\ the soft
clamp fiber bundle model where the closer two fibers are, the more they
interact.  

Hence, when we in the renormalization group scheme to be presented choose to
replace pairs of fibers by a single fiber by going from $N$ to $N/2$ 
fibers, we may choose to group them together as we like.  We explore this property in
Section \ref{order} where fibers are grouped together in terms of increasing
strength $t$.  This vastly simplifies the construction of the renormalization
group scheme. 

However, if the renormalization group scheme we present is to have any
bearing on the more complex fiber bundle models such as the LLS and the
SC models where the relative position of the fibers {\it do\/} matter,
the renormalization group presented in Section \ref{order} is
useless.  Hence, in Section \ref{real}, we present the real space
renormalization group scheme.  We map out the flow in parameter space and 
the fixed point structure.  

In section \ref{strength} we consider how the 
strength of the fiber bundle evolves under the renormalization group scheme. 

The last section \ref{conclusion} contains a discussion of our results.

\section{Fiber bundle model in a damage mechanics formulation}
\label{cdm}

We will in this section formulate the equal load sharing model in a damage mechanics
formulation based on energetic considerations.  Damage mechanics is an approach to fracture
in the continuum limit where the fractures are represented by a continuous damage parameter.
Abaimov \cite{a15} and Berthier \cite{b15} present some damage mechanics 
formulations of 
the ELS fiber bundle model.  Our approach is different from both of them.

When the fiber bundle is loaded, the fibers fail according to their thresholds, the weaker
before the stronger.  We suppose that $n$ fibers have failed.  At a load $\Delta$, the
fiber bundle carries a force
\begin{equation}
\label{eq3}
F=N\kappa (1-d)\Delta\;,
\end{equation}
where we have defined the {\it damage\/}
\begin{equation}
\label{eq4}
d=\frac{n}{N}\;
\end{equation}
and used equation (\ref{eq2-0}).
The damage parameter $d$ becomes continuous as $N\to\infty$.

A fundamental equation in what follows is the relation between damage $d$ and the
threshold of the weakest surviving fiber, $\tau$.  Since the fibers fail in a sequence
ordered from the weakest to the strongest, we have that
\begin{equation}
\label{eq5}
d=P(\tau)\;,
\end{equation}
where the cumulative probability distribution corresponding to the threshold distribution $p(t)$
is given by
\begin{equation}
\label{eq2}
P(t)=\int_0^t\ dt' p(t')\;.
\end{equation}
We will assume first that the load $\Delta$ is our control parameter. 
 Afterwards, we will
assume that the force carried by the fiber bundle $F$ is our control parameter.
We now construct the energy budget according to damage mechanics.  At a load
$\Delta$ and damage $d$, the elastic energy stored by the surviving fibers is
\begin{equation}
\label{eq6}
E^e(\Delta,d)=\frac{N\kappa}{2}\ \Delta^2\left(1-d\right)\;.
\end{equation}
We are here assuming the limit $N\to\infty$ and $\kappa\to 0$ so that
the product $N\kappa$ remains constant. 

The energy dissipated by the failed fibers is given by
\begin{equation}
\label{eq7}
E^d(d)=\frac{N\kappa}{2}\ \int_0^d\ d\delta \left[P^{-1}(\delta)\right]^2\;.
\end{equation}

We add $E^e$ and $E^d$ to get the work performed on the system to reach the state $(\Delta,d)$
from the state $(0,0)$, 
\begin{eqnarray}
\label{eq8}
W(\Delta,d)&=&E^{e}(\Delta,d)+E^{d}(d)\nonumber\\
&=&\frac{N\kappa}{2}\left[\Delta^2(1-d)+
\int_0^d\ d\delta \left[P^{-1}(\delta)\right]^2\right]\;.\nonumber\\
\end{eqnarray}

The force conjugated to the load $\Delta$ is
\begin{equation}
\label{eq9}
F=\ \left(\frac{\partial W}{\partial \Delta}\right)_d=N\kappa \Delta (1-d)\;,
\end{equation}
which is identical equation (\ref{eq3}), as it must.

The damage driving force $\cal F$ conjugate to the damage $d$ is
\begin{equation}
\label{eq10}
{\cal F} =-\ \left(\frac{\partial W}{\partial d}\right)_\Delta= \frac{N\kappa}{2}
\left[\Delta^2-\left[P^{-1}(d)\right]^2\right]\;,
\end{equation}
and the equilibrium condition is
\begin{equation}
\label{eq11}
{\cal F}=0\;,
\end{equation}
which when combined with equation (\ref{eq10}) gives
\begin{equation}
\label{eq12}
\Delta=P^{-1}(d)\;.
\end{equation}
This equation is equivalent to equation (\ref{eq5}) when $t=\Delta$ and it simply
states that at a load $\Delta$ all fibers with threshold less than or equal to it
have failed.

We now turn to controlling the force $F$ rather than the load $\Delta$.  Using
equation (\ref{eq3}), we have
\begin{equation}
\label{eq13}
\Delta(F)=\frac{F}{N\kappa}\ \frac{1}{1-d}\;.
\end{equation}
The corresponding work we find via the Legendre transform,
\begin{equation}
\label{eq14}
U(F,d)=W\left(\Delta(F),d\right)-F\Delta(F)\;.
\end{equation}
Combining this equation with equations (\ref{eq8}) and (\ref{eq13}) we find
\begin{equation}
\label{eq15}
U(F,d)=-\ \frac{F^2}{2N\kappa}\ \frac{1}{1-d}+\frac{N\kappa}{2}\
\int_0^d\ d\delta \left[P^{-1}(\delta)\right]^2\;.
\end{equation}
We calculate the damage driving force
\begin{equation}
\label{eq16}
{\cal F}= -\ \left(\frac{\partial U}{\partial d}\right)_F = \frac{F^2}{2N\kappa}\ \frac{1}{(1-d)^2}
-\ \frac{N\kappa}{2}\ \left[P^{-1}(d)\right]^2\;.
\end{equation}
The equilibrium condition (\ref{eq11}) gives
\begin{equation}
\label{eq17}
F=N\kappa(1-d)P^{-1}(d)\;.
\end{equation}

Equation (\ref{eq17}) when combined with Equation (\ref{eq5}) gives
\begin{equation}
\label{eq18}
F=N\kappa\left[1-P\left(\Delta(F)\right)\right]\Delta(F)\;,
\end{equation}
which is the force-load characteristics of the fiber bundle model.  This equation is
usually derived using order statistics.  We see that the derivation using 
damage mechanics leads to the same result.

It is interesting to note that the equilibrium condition (\ref{eq11}) can only be
satisfied for
\begin{equation}
\label{eq19}
\frac{F}{N\kappa} \le \max_d (1-d)P^{-1}(d)\;.
\end{equation}
If $F$ exceeds this limit, $\cal F$ is positive and catastrophic failure ensues.

\section{Renormalization group}
\label{rg}

The renormalization group transformation that we are about to construct will
consist of replacing the original fiber bundle containing $N$ individual fibers
by a new fiber bundle containing $N/2$ individual fibers.  We introduced the
damage parameter $d=n/N$ in equation (\ref{eq4}), where $n$ is the number of
failed fibers.  It is only in the limit $N\to\infty$ that $d$ is a continuous 
parameter.  It is convenient in the following to use the notation $d(n,N)=n/N$ 
in order to indicate that for finite $N$, $d(n,N)$ is a discrete variable.  
We note the following equality,
\begin{equation}
\label{eq20}
d\left(n,N\right)=d\left(\frac{n}{2},\frac{N}{2}\right)\;.
\end{equation} 
We now demand that the total work performed on the system,  (\ref{eq8}), is kept constant by the 
renormalization group transformation.  That is, we have
\begin{equation}
\label{eq21}
W_N\left(\Delta,d(n,N)\right)=W_{N/2}\left(\Delta,d\left(\frac{n}{2},\frac{N}{2}\right)\right)\;.
\end{equation}
As we are here assuming $N$ to be finite, we have explicitly introduced it as a parameter in writing
$W\to W_N$.  
Equation (\ref{eq21}) is central in what follows.  As the number of individual fibers is
reduced from $N$ to $N/2$, the possible values that the damage parameter $d(n,N)$
can take is also reduced by a factor 2.  However, {\it for those values of the damage
parameter that remain, the energy is unchanged.\/}  

The renormalization group transform consists of replacing pairs of fibers in the original
fiber bundle by single fibers.  We have just stated that the energy (\ref{eq8}) is to remain
constant under the transformation.  The energy consists of two parts, $W_N=E^e_N+E^d_N$, see equations
(\ref{eq6}) and (\ref{eq7}). In order to fulfill equation (\ref{eq21}), the elastic energy
$E^e_N$ and the energy dissipated by the damage $E^d_N$ each needs to be constant under the 
renormalization group transition.    

In order for the elastic energy to be constant under the renormalization group transformation,
we need to transform the elastic constant.  The elastic energy will be constant if keep the load $\Delta$ 
fixed, i.e.,
\begin{equation}
\label{eq21-0}
\Delta\to\Delta'=\Delta\;,
\end{equation}
and we set
\begin{equation}
\label{eq22}
\kappa\to\kappa'=2\kappa
\end{equation}
so that
\begin{equation}
\label{eq23}
N\kappa=\left[\frac{N}{2}\right]\ \left[2 \kappa\right]\;,
\end{equation}
when
\begin{equation}
\label{eq22-1} 
N\to N'=\frac{N}{2}\;.
\end{equation}

We keep the load $\Delta$ fixed during the transformation.  

The energy dissipated by the failed fibers, $E^d_N$, is constant under the renormalization 
group transformation if, in the limit $N\to\infty$, we have 
\begin{eqnarray}
\label{eq24}
E^d_N&=&\frac{N\kappa}{2}\ \int_0^{d(n,N)}\ d\delta \left[P^{-1}(\delta)\right]^2\nonumber\\
&=&\frac{(N/2)(2\kappa)}{2}\ \int_0^{d(n/2,N/2)}\ d\delta \left[P^{-1}(\delta)\right]^2\;.\nonumber\\
\end{eqnarray}

When $N$ is finite, equation (\ref{eq24}) becomes
\begin{equation}
\label{eq25}
\frac{\kappa}{2}\ \sum_{i=1}^n t_{(i)}^2=\frac{2\kappa}{2}\ \sum_{j=1}^{n/2} {t'}_{(j)}^2\;,
\end{equation}
where we have ordered the thresholds, $t_{(1)}\le t_{(2)} \le \dots \le t_{(N-1)}\le t_{(N)}$ and
${t'}_{(1)}\le {t'}_{(2)} \le \dots \le {t'}_{(N/2-1)}\le {t'}_{(N/2)}$.  

These ordered thresholds are averaged over an ensemble.  That is, we have $M$ samples.  The $k$th 
largest threshold in sample $m$ is $t_{(k)}^m$, and we have
\begin{equation}
\label{eq25-0}
t_{(k)}=\frac{1}{M}\ \sum_{m=1}^M\ t_{(k)}^m\;.
\end{equation}
    
A fundamental result in order statistics is that the averaged $k$th ordered threshold is
given by
\begin{equation}
\label{eq25-1}
P\left(t_{(k)}\right)=\frac{k}{N+1}\;,
\end{equation}
in the limit when $M\to\infty$, see Gumbel \cite{g04}. 

In order to complete the renormalization group transformation, we need to define the threshold
transformation
\begin{eqnarray}
\label{eq26}
& &\left\{t_{(1)},t_{(2)},\dots,t_{(N-1)},t_{(N)}\right\}\to\nonumber\\
& &\left\{{t'}_{(1)},{t'}_{(2)},\dots,{t'}_{(N/2-1)},{t'}_{(N/2)}\right\}\nonumber\\
\end{eqnarray}
so that equation (\ref{eq25}) is fulfilled.  There is no unique way to do this.  We will in the 
following present two different transformations. The first one, which we call the {\it order space\/}
transformation, consists of grouping the initial thresholds according to their value.  The second
one, the {\it real space\/} transformation, consists of grouping the initial thresholds according to 
their location.      

\subsection{Renormalization group in order space}
\label{order}

We write the sum in (\ref{eq25}) as
\begin{equation}
\label{eq27}
\sum_{i=1}^n\ t_{(i)}^2= 2\ \sum_{j=1}^{n/2}\ \left[\frac{t_{(2j-1)}^2+t_{(2j)}^2}{2}\right]
=2\ \sum_{j=1}^{n/2}\ {t'}_{(j)}^2\;.
\end{equation}
Hence, we define the {\it order space\/} threshold transformation {\it at the indivdual sample
level\/} as
\begin{equation}
\label{eq28}
t_{(i)}^m\to {t'}_{(j)}^m=\left[\frac{(t_{(2j-1)}^m)^2+(t_{(2j)}^m)^2}{2}\right]^{1/2}\;,
\end{equation}
where as in equation (\ref{eq25-0}) the index $m$ identifies the sample.

Equations (\ref{eq21-0}), (\ref{eq22}), (\ref{eq22-1}) and (\ref{eq28}) define 
the order space renormalization group transformation, fulfilling equation (\ref{eq21}).

\begin{figure} 
\begin{center}  
\includegraphics[width=7cm]{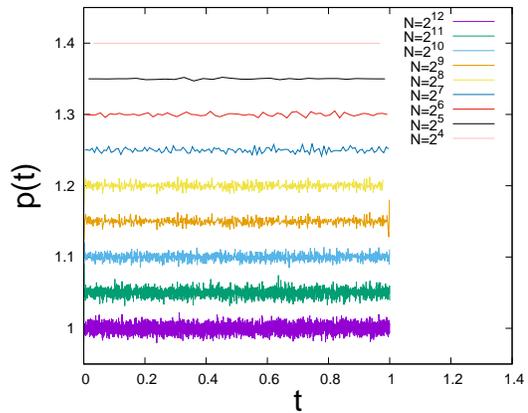} 
\caption{\label{fig1} 
Evolution of threshold distribution for $M=10^6$ fiber bundles, each
containing $N=2^{12}$ fibers when repeating the order space renormalization
group transformation.  The initial threshold distribution was uniform on the
unit interval. We have added a constant factor to $p(t)$ for each iteration 
of the renormalization group in order to separate them.} 
\end{center} 
\end{figure} 

Using equation (\ref{eq25-1}), we have
\begin{equation}
\label{eq29}
P(t_{(2j-1)})=P(t_{(2j)})-\frac{1}{N}\;,
\end{equation}
when $N\gg 1$.  To first order in $1/N$, this gives
\begin{equation}
\label{eq30}
t_{(2j-1)}=t_{(2j)}-\frac{1}{Np(t_{(2j)})}\;.
\end{equation} 
Combining this expression with renormalization group transformation (\ref{eq28}), gives
\begin{equation}
\label{eq31}
{t'}_{(j)}=t_{(2j)}-\frac{1}{2Np(t_{(2j)})}\;,
\end{equation}
where ${t'}_{(j)}$ is the $j$th smallest average threshold out of $N/2$ and $t_{(2j)}$ 
is the $2j$th average threshold out of $N$.  Hence, for large $N$, any threshold 
distribution $p(t)$ will be invariant under this renormalization group transformation.

We show in figure \ref{fig1}, the evolution of the distribution of individual thresholds 
of $M=10^6$ samples, each having $N=2^{12}$ fibers, as we reiterate the order space 
renormalization group transformation.  The $N=2^{12}$ thresholds for the initial system
were generated from a flat distribution on the unit interval.  As we see from the figure, the
distribution does not change as the renormalization group transformation is iterated.    

The renormalization group transformation in terms of the parameters of the model, (\ref{eq21-0}), 
(\ref{eq22}), and (\ref{eq28}), defines the flow in parameters space,
\begin{equation}
\label{eq31-1}
\left(\begin{array}{c}
\Delta\\      
\kappa\\
d\\
\end{array}\right)
\to
\left(\begin{array}{c}
\Delta'\\      
\kappa'\\
d'\\
\end{array}\right)\;.
\end{equation}
We will now study the flow of the parameters $(\Delta,\kappa,d)$ under the 
renormalization group.

\begin{figure}
\begin{center} 
\includegraphics[width=7cm]{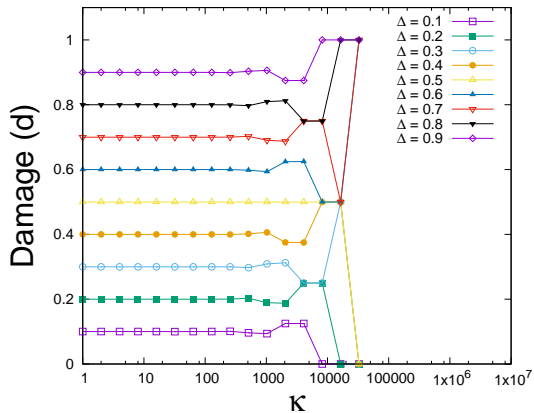} 
\caption {\label{fig2} Flow in parameter space under the renormalization group
transformation (\ref{eq31-1}) projected onto the $(\kappa,d)$ plane. 
The initial number of fibers was $N=2^{15}$, and the
data are averaged over $10^6$ samples.  The renormalization group is iterated 15 times so
that the last bundle contains one fiber.  The threshold distribution was uniform on the
unit interval.} 
\end{center} 
\end{figure} 

For the uniform distribution on the unit interval, i.e.\ $P(t)= t$, the work 
(\ref{eq8}) is
\begin{equation}
\label{eq31-2}
W_N(\Delta,d)=\frac{N\kappa}{2}\left[\Delta^2(1-d)+\frac{d^3}{3}\right]\;.
\end{equation} 
We will in the following assume this threshold distribution for simplicity. 

\begin{figure}[htb]
\begin{center} 
\includegraphics[width=7cm]{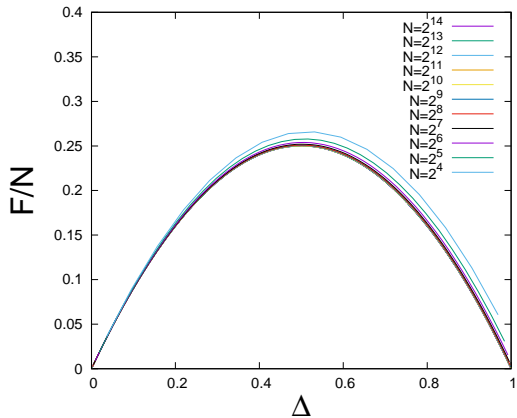}  
\caption {\label{fig3} 
The force-load curve as we iterate the order-space renormalization
group. The initial number of fibers was $N=2^{15}$ and the uniform
distribution on the unit interval was assumed. Averages were taken over 
$10^6$ samples.} 
\end{center} 
\end{figure}

We assume that $N$ is finite.  Let us define a strain $\Delta_N(n)$ such that
if $\Delta>\Delta_N(n)$, at least $n$ fibers have failed when the bundle
is in equilibrium whereas if $\Delta<\Delta_N(n)$, up to $n-1$ fiber have
failed when the bundle is in equilibrium. We calculate the value of $\Delta_N(n)$
by demanding continuity, 
\begin{eqnarray}
\label{eq31-06}
&&\lim_{\delta\to 0}W_N\left(\Delta_N(n)+\delta,\frac{n}{N}\right)\nonumber\\
&=&\lim_{\delta\to 0}W_N\left(\Delta_N(n)-\delta,\frac{n-1}{N}\right)\;,\nonumber\\
\end{eqnarray}  
which for the uniform distribution on the unit interval gives 
\begin{equation}
\label{eq31-08}
\Delta_N(n)=\frac{\sqrt{1-3n+3n^2}}{\sqrt{3}N}\;.
\end{equation}
For a given load $\Delta$, we then have
\begin{equation}
\label{eq31-100}
\dots<\Delta_N(n-1)\le\Delta\le\Delta_N(n)<\Delta_N(n+1)<\dots\;,
\end{equation}
giving rise to the flow diagram shown in figure \ref{fig2}, which is 
a projection into the $(\kappa,d)$ plane of the flow (\ref{eq31-1}). 
For each iteration where $N\to N/2$ and $\kappa\to 2\kappa$, the position 
of $\Delta$ in the sequence
of inequalities (\ref{eq31-100}) determines the damage level.   

\begin{figure}[htb]
\begin{center}  
\includegraphics[width=7cm]{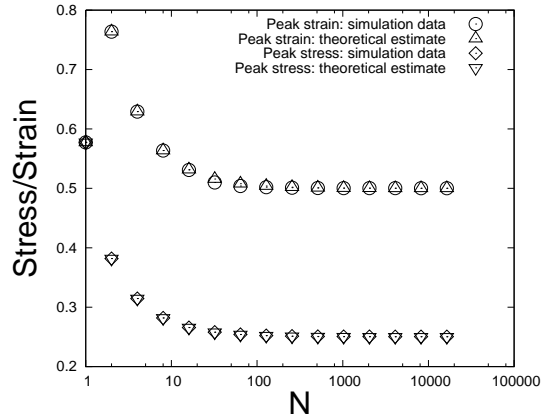} 
\caption {\label{fig4} $\Delta_N^c$ and the corresponding force $F_N^c/N$
as we iterate the order-space renormalization group. The initial number of fibers 
was $N=2^{15}$ and the uniform distribution on the unit interval was assumed. 
Averages were taken over $10^6$ samples.} 
\end{center} 
\end{figure}

We show in figure \ref{fig3} the force-load curve for each 
renormalization group iteration.  For each iteration, there is 
load $\Delta_N^c$ for which the force $F$ is maximal.  This maximum
occurs for
\begin{equation}
\label{eq31-07}
n=\left\{ \begin{array}{ll}
1 & \mbox{if $N=1$}\;,\\
\frac{N}{2}+1 & \mbox{if $N=2,\ 4,\cdots$}\;.\\
                  \end{array}
\right.
\end{equation}
Combined with equation (\ref{eq31-08}), this gives 
\begin{equation}
\label{eq31-10}
\Delta_N^c=\left\{ \begin{array}{ll}
\frac{1}{\sqrt{3}} & \mbox{if $N=1$}\;,\\
\frac{1}{2N}\ \left[\frac{4}{3}+N\left(N+2\right)\right]^{1/2} & \mbox{if $N=2,\ 4,\cdots$}\;,\\
                  \end{array}
\right.
\end{equation} 
and the corresponding peak stress is 
\begin{equation}
\label{eq31-12}
\frac{F_N^c}{N}=\left\{ \begin{array}{ll}
\kappa\Delta_N^c & \mbox{if $N=1$}\;,\\
\frac{\kappa\Delta_N^c}{2} & \mbox{if $N=2,\ 4,\cdots$}\;.\\
                  \end{array}
\right.
\end{equation}   
This is illustrated in figure \ref{fig4}.

\subsection{Renormalization group in real space}
\label{real}

The order space renormalization group we have defined and explored in section \ref{order} is tailored for the ELS
fiber bundle model since the physical position of the fibers do not matter.  Hence, for the renormalization group 
procedure to be generalizable to more complex models than the ELS fiber bundle model, the LLS fiber bundle model being 
an example, we group {\it neighboring\/} fibers together.  

We assume the fibers to be placed along a one-dimensional line.  They are numbered from 1 to $N$.  Hence, we are considering
a one-dimensional system.  The procedure that we describe is straight forward to generalize to e.g.\ having the fibers positioned
at the nodes of a square lattice. 

\begin{figure}
\begin{center} 
\includegraphics[width=7cm]{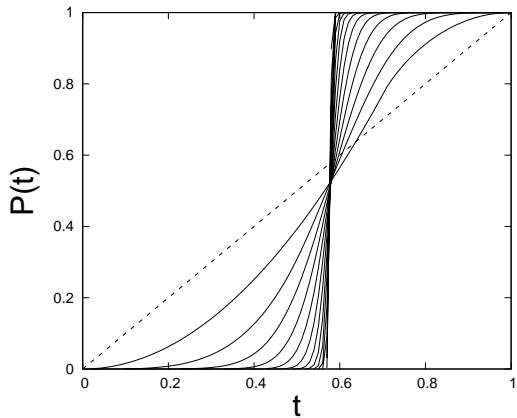} 
\caption {\label{fig5} 
Evolution of the cumulative threshold probability under the 
real space renormalization group iteration is shown. 
The initial number of fibers was $N=2^{15}$. A uniform threshold 
distribution on the unit interval is assumed, so that the
cumulative probability is $P(t)=t$ (dotted line) at the initial stage.
 Averages are 
taken over $10^6$ samples.} 
\end{center} 
\end{figure}

We follow the same procedure as for the order space renormalization group except that the group together of pairs of fibers
are now in real space rather than in order space.  Fiber number $i$ has a threshold $t_i$.  The renormalization group transformation
of the thresholds then becomes
\begin{equation}
\label{eq28-1}
t_{(i)}^m\to {t'}_{j}^m=\left[\frac{(t_{2j-1}^m)^2+(t_{2j}^m)^2}{2}\right]^{1/2}\;
\end{equation}
at the individual sample level. As in the order space renormalization group, the work performed on
the fiber bundle is conserved, see equation (\ref{eq21}).  For an $N=2$ fiber bundle, the order and real space renormalization
groups are identical.  Hence, we may see the real space renormalization group as (1) grouping neighboring fibers into fiber bundles
of size $N=2$ and (2) do an order space renormalization group iteration on each pair of fibers.

\begin{figure}
\begin{center} 
\includegraphics[width=7cm]{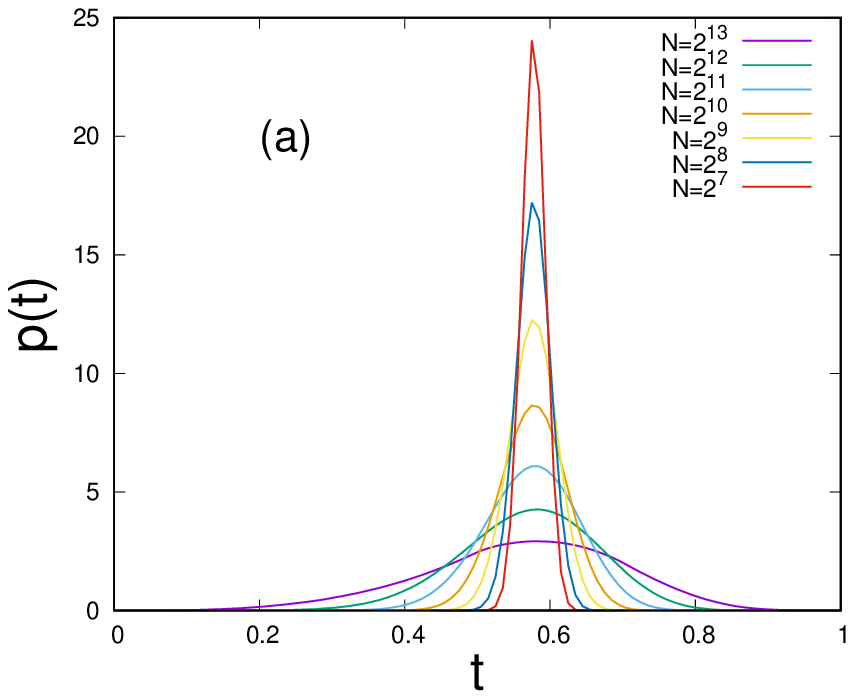} 
\includegraphics[width=7cm]{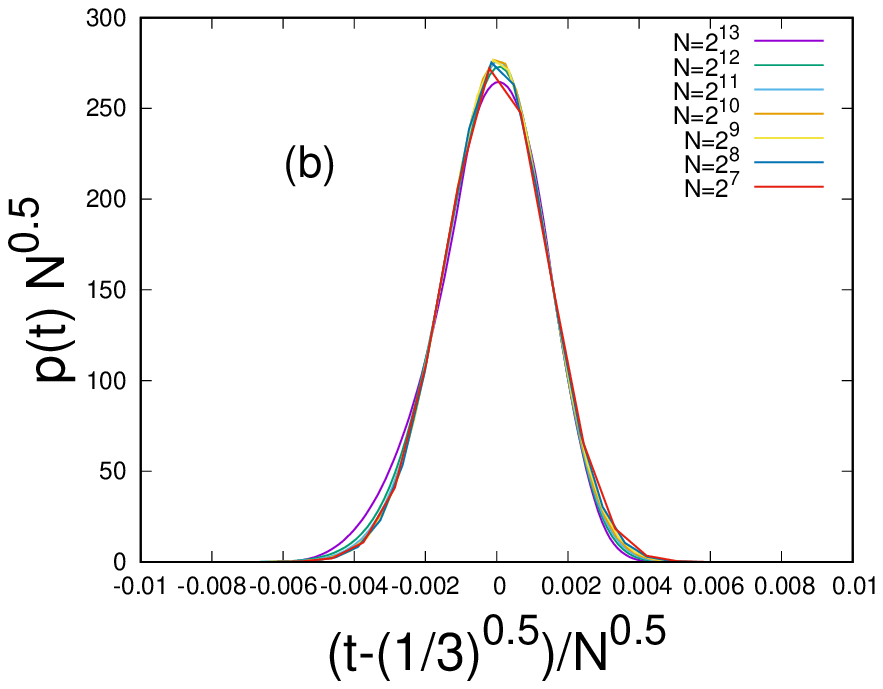} 
\caption {\label{fig6} 
(a) Evolution of the uniform threshold distributions on the unit interval under
the real space renormalization group. (b) Rescaling the thresholds leads to
a data collapse. The initial number of fibers was $N=2^{15}$.  Averages were 
taken over $10^6$ samples.} 
\end{center} 
\end{figure}

In contrast to the order space renormalization group, the threshold distribution is {\it not\/} invariant under the real
space renormalization group.  We show the evolution of the uniform distribution on the unit interval in figures \ref{fig5}
and \ref{fig6}.  In figure \ref{fig5} we show the cumulative probability $P(t)$ as it evolves from the initial $P(t)=t$. We note
that all the iterated cumulative probabilities pass through the same point $(t_c,P(t_c))$. In figure \ref{fig6}, showing the 
evolution of the threshold distribution, we see that the threshold distribution is becoming increasingly peaked around $t_c$.   
Since the work is conserved as for the order space renormalization group, we will have that $\Delta_1(1)=1/\sqrt{3}$ also in 
this case. Since the threshold distribution is no longer invariant under the renormalization group, but approaches a delta function, 
we must have that $t_c=1/\sqrt{3}$.  
We can  calculate the fixed point value $t_c=1/\sqrt{3}$ through the following 
argument, which is valid both for order and space renormalization group 
schemes: 
The final single fiber must have a strength which
can conserve the energy of the whole bundle (with $N$ fibers) we started with.
From the energy conservation, we can write for an uniform distribution of fiber 
thresholds within ($0,1$) 
\begin{equation}
\frac{1}{2}N\kappa t_c^2 =\frac{1}{2}N\kappa\int_0^1 t^2p(t)dt;
\label{eq:t_c}
\end{equation}
which gives $t_c=1/\sqrt3$.
Since $W_N(\Delta_N(n),n)$ is a monotonously increasing function in $n$, and 
therefore
also $\Delta_N(n)$, we must have that $t_c$ is a symmetry point with half the thresholds smaller and half the thresholds larger
than $t_c$.  Hence, we have $P(t_c)=1/2$.  This is what is seen in figure \ref{fig5}. 

\begin{figure}
\begin{center} 
\includegraphics[width=7cm]{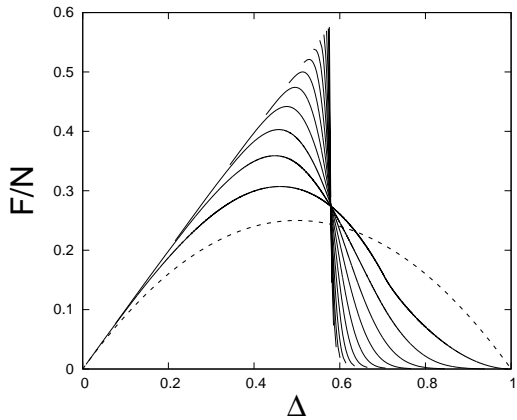} 
\caption {\label{fig7} 
Evolution of the force-load curve
under the real space renormalization group. The starting point
was $10^6$ fiber bundles each containing $N=2^{15}$ fibers. The threshold 
distribution was uniform on the unit interval. The dotted curve  $\Delta(1-\Delta)$ is the initial force-load curve.} 
\end{center} 
\end{figure}

We observe numerically that $p(t)$ assumes almost symmetric Gaussian like 
distribution peaked around the critical load value $t_c=1/\sqrt3$. The variance falls off as
$1/\sqrt{N}$. We show this in figure \ref{fig6} where the data have been fitted to the function 
\begin{equation}
\label{eq41}
p(t)=\frac{1}{\sqrt N}\ \phi\left(\frac{t-\frac{1}{\sqrt 3}}{\sqrt N}\right);
\end{equation}
where 
\begin{equation}
\label{eq42}
\phi(y)=A \exp(-By^2)\;,
\end{equation}
with $A=277$ and $B=240000$. 

\begin{figure}
\begin{center} 
\includegraphics[width=7cm]{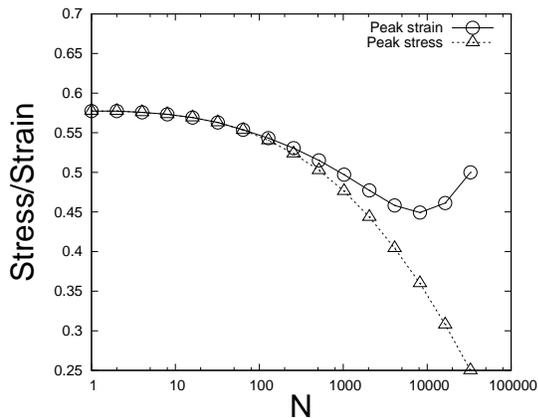} 
\caption {\label{fig8} 
Evolution of the peak-stress and peak-strain 
under the real space renormalization group. The starting point
was $10^6$ fiber bundles each containing $N=2^{15}$ fibers. The threshold 
distribution was uniform on the unit interval.} 
\end{center} 
\end{figure}

We show in figure \ref{fig7} the evolution of the force-load curve under
the real space renormalization group, i.e., the equivalent of figure \ref{fig3}
for the order space renormalization group. For a given cumulative threshold
probability $P(t)$, the force-load curve will be given by equation (\ref{eq18}).
Hence, the curves seen in figure \ref{fig7} reflect this equation combined with
equation (\ref{eq41}) giving the evolution of the threshold distribution.
Figure \ref{fig8} shows how the evolution of threshold distribution influences 
the peak-stress and peak-strain values. After few renormalization steps, 
peak-stress and peak-strain values converge to same level.     

\begin{figure}
\begin{center} 
\includegraphics[width=7cm]{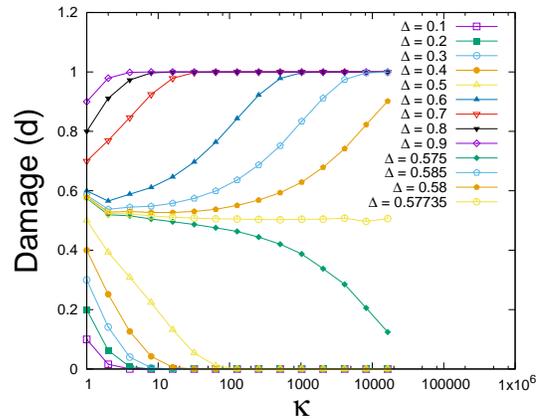} 
\caption{\label{fig9} The flow in $(\Delta,\kappa,d)$ space
projected into the $(\kappa,d)$ plane, under the real space renormalization
group. Initial number of fibers $N=2^{15}$ having uniform fiber strength 
distribution on the unit interval. Averages were taken over $10^3$ samples.} 
\end{center} 
\end{figure} 

The damage parameter $d$ is directly linked to the threshold distribution of the 
fibers. Under the real space renormalization group scheme, the threshold distribution 
is not invariant. Hence, $d$ evolves throughout the entire renormalization group
iteration and not just towards the end of the process as in the order space
renormalization group scheme, see figure \ref{fig2}.  We have shown in figure \ref{fig9}
the equivalent flow diagram for the real space renormalization group.  The flow
appears in $(\Delta,\kappa,d)$ space but has been projected into the 
$(\kappa,d)$ plane, see equation (\ref{eq31-1}).  There are three fixed points:
$d=0$, $d=d_c$ and $d=1$.  The first and the last are stable and $d_c$ is
unstable.  For the uniform threshold distribution on the unit interval, we have
$d_c=1/2$. 

\begin{figure}
\begin{center}
\includegraphics[width=7cm]{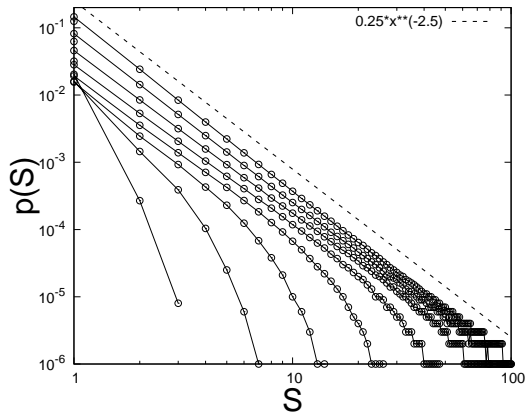}
\caption {\label{fig10} Evolution of the histogram of avalanche sizes
as the real space renormalization group is iterated.  The initial threshold
distribution was uniform on the unit interval. 
The initial number of fibers was $N=2^{14}$ and averages were taken over $10^5$ samples.}
\end{center}
\end{figure}

An important aspect of the renormalization group is how fluctuations are
handled.  Here we consider the avalanche distribution \cite{hh92,phc10,hhp15}.
The size of an avalanche is the number of fibers that fail simultaneously
as a result of a change of the external parameters.  We consider here
a change in the external force $F$ under quasi-static conditions. It was shown under 
very general conditions that the histogram would follow a power law with exponent 
$-5/2$ \cite{hh92}.  We show in figure \ref{fig10} the evolution of the avalanche 
histogram under the real space renormalization group.  The power law character of 
the avalanche distribution remains until the number of fibers in the bundle is
too low.  The exponent $-5/2$ remain in place as long as there is a discernable 
power law.         

\begin{figure}
\begin{center}
\includegraphics[width=7cm]{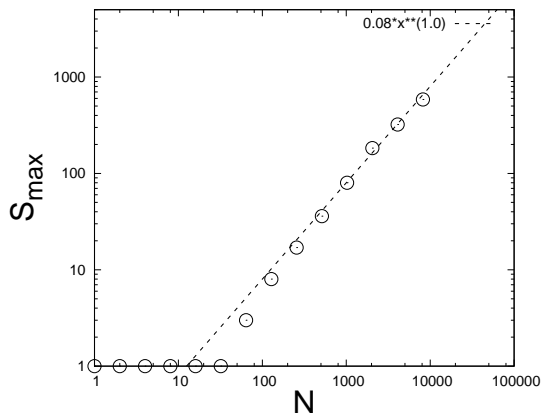}
\caption {\label{fig11}
Evolution of the largest avalanche occurring before complete failure
as the real space renormalization group is iterated. 
The initial threshold distribution was uniform on the unit interval. 
The initial number of fibers was $N=2^{14}$ and averages were taken over 
$10^5$ samples.}
\end{center}
\end{figure}

We plot in figure \ref{fig11} the average of the largest avalanche occurring before 
complete failure of the bundle as a function of bundle size $N$ under the real space 
renormalization group. We see that this average reaches unity --- the smallest value
it can take on --- when the fiber bundle is reduced to a bundle of $2^5$ fibers 
(we have started with a bundle of $2^{14}$ fibers).  

\section{Renormalized strength}
\label{strength}
\begin{figure}
\begin{center} 
\includegraphics[width=7cm]{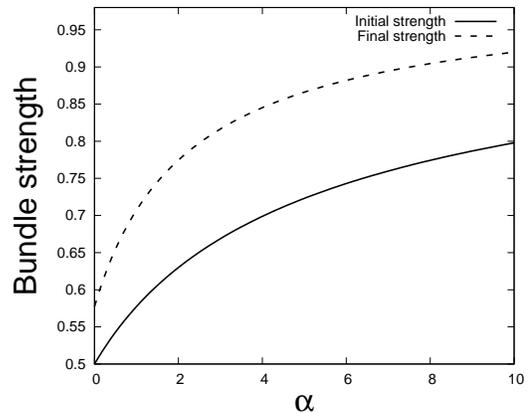} 
\caption {\label{fig12} 
Initial and final strengths of the fiber bundle vs.\ power law 
index ($\alpha$) under both the order space and the real space renormalization group
schemes. For $\alpha=0$, the distribution is reduced to 
the uniform distribution and the initial and final strength values match well 
with our numerical results.} 
\end{center} 
\end{figure}

Now we are going to compare the {\it initial} and {\it final} strengths of a 
fiber bundle, which has gone through the renormalization group scheme. 
We have, so far only considered the uniform threshold distribution on the unit 
interval.
In the following, we consider a power law on the unit interval,
\begin{equation}
\label{eq43}
p(t)=(1+\alpha)t^\alpha,
\end{equation}
where $\alpha \ge  0$. When $\alpha=0$, we have the uniform distribution. 
We will study 
the initial and final fiber bundle strength at imposed load $\Delta$ under the order or real
space renormalization scheme. The force on a bundle at load $\Delta$ is 
\begin{equation}
\label{eq44}
F(\Delta)=N\kappa\Delta(1-P(\Delta))=N\kappa\Delta(1-\Delta^{\alpha+1});.
\end{equation}
By solving $dF/d\Delta=0$ for $\Delta$ gives us the initial strength of the 
fiber bundle (with $N$ fibers),
\begin{equation}
\label{eq45}
(\Delta^c_N)_{initial}=(\frac{1}{\alpha+2})^\frac{1}{1+\alpha}\;.
\end{equation}
Since the work performed on the fiber bundle is conserved by the 
renormalization
group, at the final renormalization step (when $N=1$) we must 
have (see Eq. \ref{eq:t_c}) 
\begin{equation}
(\Delta^c_N)_{final}^2 = \int_0^1 t^2p(t)dt;
\label{eq46}
\end{equation}
which gives the final strength of the bundle (with $N=1$ fiber)  
\begin{equation}
(\Delta^c_N)_{final} =\sqrt\frac{\alpha+1}{\alpha+3}\;.  
\label{eq47}
\end{equation}
Figure \ref{fig12} compares the initial and final bundle strengths as a function of 
power law index $\alpha$. 

We see that the final strength of a bundle has gone up as a result of the 
renormalization group scheme !

\section{Conclusion and discussion}
\label{conclusion}

We have in this paper introduced a renormalization group for the equal 
load sharing fiber bundle model based on formulating the fiber bundle 
in the context of damage mechanics.  The idea behind the damage mechanics
formulation is to introduce a continuous damage variable so that the
binary nature of the single fibers is no longer in focus.  In this way, we
are able to group together the fibers belonging to a given fiber bundle 
into smaller fiber bundles and map the parameters of the larger
fiber bundle onto the smaller bundles.  A central concept in this mapping
is the conservation of the work applied to the fiber bundle to create a
certain level of damage, equation (\ref{eq8}).  This work is kept invariant
under the renormalization group procedure.  

We have presented two versions of the renormalization group.  In the 
{\it order space\/} formulation, we group the fibers together according to
their failure strength. The three parameters $(\Delta,\kappa,d)$ are mapped
according to equations (\ref{eq21-0}), (\ref{eq22}) and (\ref{eq28}).  
Under this version of the renormalization group, the threshold distribution
remains invariant.  

The problem with the order space renormalization group scheme is that
there is no obvious way to generalize it to other fiber bundle models 
such as the local load sharing model.  Rather than grouping together
the fibers according to their strength, we may group them together 
according to their locations, hence defining the {\it real space\/}
renormalization group scheme.  The flow equations are then given by
(\ref{eq21-0}), (\ref{eq22}) and (\ref{eq28-1}).  

We have in this paper only presented the renormalization group schemes
themselves together with a number of their properties.  We have not
attempted to implement the renormalization group on more complex fiber bundle
 models. 
We do see a strong potential in the use of the renormalization group
as a tool to investigate the fiber bundle models, in particular in
connection with fluctuations (see figure \ref{fig1}) and strength enhancement (see figure \ref{fig12}).            

\begin{acknowledgments}
The authors thank Martin Hendrick, Jonas T.\ Kjellstadli and Laurent Ponson
for interesting discussions.  This work was partly supported by the
Research Council of Norway through its Centers of Excellence funding
scheme, project number 262644.
\end{acknowledgments}



\end{document}